\documentclass[]{aa}
\usepackage{graphicx,natbib}

\pdfoutput=1

\def\jj{\mbox{$J$}}

\def\ks{\mbox{$K_s$}}

\def\jk{\mbox{$(J-K_s)$}}

\def\aV{\mbox{$A_{V}$}}
\def\dV{\mbox{$\delta A_V$}}
\def\mdV{\mbox{$\overline{\delta A}_V$}}
\def\sdr{\mbox{$\sigma (\dV)$}}
\def\mMJ{\mbox{$(m-M)_J$}}
\def\eJK{\mbox{$E(J-K_s)$}}
\def\ms{\mbox{$\rm M_\odot$}}
\def\ds{\mbox{$d_\odot$}}

\def\ta{\mbox{$t_{clu}$}}
\def\ts{\mbox{$t_*$}}
\def\sms{\mbox{$m_*$}}
\def\Mcl{\mbox{$M_{clu}$}}
\def\x2{\mbox{$R_{rms}$}}
\def\fB{\mbox{$f_{bin}$}}
\def\nS{\mbox{$N_{sim}$}}
\def\nR{\mbox{$N_{run}$}}
\def\sfs{\mbox{$\tau_{sfs}$}}
\def\rr{\mbox{$ASA_{min}$}}
\def\P{\mbox{$\vec{P}$}}
\def\Pm{\mbox{$\vec{P}_{min}$}}

\begin{document}

\title{Unveiling hidden properties of young star clusters: differential 
reddening, star-formation spread and binary fraction}

\author{C. Bonatto\inst{1} \and E.F. Lima\inst{1} \and E. Bica\inst{1}}

\offprints{C. Bonatto}

\institute{Universidade Federal do Rio Grande do Sul, Departamento de Astronomia\\
CP\,15051, RS, Porto Alegre 91501-970, Brazil\\
\email{charles.bonatto@ufrgs.br, eliade.lima@ufrgs.br, bica@if.ufrgs.br}
\mail{charles.bonatto@ufrgs.br} }

\date{Received --; accepted --}

\abstract
{Usually, important parameters of young, low-mass star clusters are very difficult 
to obtain by means of photometry, especially when differential reddening and/or 
binaries occur in large amounts.}
{We present a semi-analytical approach (\rr) that, applied to the Hess diagram of 
a young star cluster, is able to retrieve the values of mass, age, star-formation 
spread, distance modulus, foreground and differential reddening, and binary fraction.}
{The global optimisation method known as adaptive simulated annealing (ASA) is used 
to minimise the residuals between the observed and simulated Hess diagrams of a star
cluster. The simulations are realistic and take the most relevant parameters of young 
clusters into account. Important features of the simulations are: a normal (Gaussian)
differential reddening distribution, a time-decreasing star-formation rate, the unresolved 
binaries, and the smearing effect produced by photometric uncertainties on Hess diagrams. 
Free parameters are: cluster mass, age, distance modulus, star-formation spread, foreground 
and differential reddening, and binary fraction.}
{Tests with model clusters built with parameters spanning a broad range of values show that 
\rr\ retrieves the input values with a high precision for cluster mass, distance modulus and
foreground reddening, but somewhat lower for the remaining parameters. Given the statistical
nature of the simulations, several runs should be performed to obtain significant convergence 
patterns. Specifically, we find that the retrieved (absolute minimum) parameters converge to 
mean values with a low dispersion as the Hess residuals decrease. When applied to actual 
young clusters, the retrieved parameters follow convergence patterns similar to the models. 
We show how the stochasticity associated with the early phases may affect the results, 
especially in low-mass clusters. This effect can be minimised by averaging out several 
twin clusters in the simulated Hess diagrams.}
{Even for low-mass star clusters, \rr\ is sensitive to the values of cluster mass, age, 
distance modulus, star-formation spread, foreground and differential reddening and, to a 
lesser degree, binary fraction. Compared with simpler approaches, the inclusion of binaries, 
a decaying star-formation rate and a normally distributed differential reddening, appear 
to yield more constrained parameters, especially the mass, age and distance from the Sun.
A robust determination of cluster parameters may have a positive impact on many fields. For
instance, age, mass and binary fraction are important for establishing the dynamical 
state of a cluster, or deriving a more precise star-formation rate in the Galaxy.}

\keywords{({\it Galaxy}:) open clusters and associations; {\it Galaxy}: structure}

\titlerunning{Hidden properties of young star clusters}

\maketitle

\section{Introduction}
\label{Intro}

Several events occurring on short timescales combine to make the first few $10^7$\,yr 
the most critical period in a star cluster's life, especially for the poorly-populated 
embedded clusters (ECs). Driven mainly by the impulsive removal of the parental molecular 
gas by supernovae and massive-star winds, the rapid and deep changes in the cluster 
internal dynamics lead to the escape of varying fractions of member stars to the field. 
The end result is that most of the low-mass clusters are dissolved before reaching 
$\sim40$\,Myr of age (e.g. \citealt{tutu78}; \citealt{GoBa06}). This scenario is 
consistent with current estimates suggesting that less than $\sim5\%$ of the Galactic 
ECs dynamically evolve into gravitationally bound open clusters (e.g. \citealt{LL2003}; 
\citealt{SFR}).

Low-mass clusters undergoing such a rapidly changing phase are usually characterised
by an under-populated main sequence (MS) and significant fractions of pre-main 
sequence (PMS) stars, all mixed up with a spatially non-uniform dust and gas distribution.
To complicate matters, star formation within a cluster appears to extend for a period 
that may be comparable to the cluster age (e.g. \citealt{Stauffer97}), while the MS and 
PMS stars may be arranged in (unknown) fractions of binaries (e.g. \citealt{Weidner09}). 
Besides, the vast majority of the binary pairs cannot be resolved in Colour-Magnitude 
Diagrams (CMDs). As a result, the CMDs of young, low-mass star clusters tend to be very 
difficult to interpret. And, consequently, the determination of fundamental cluster 
parameters (e.g. age, distance, mass, foreground and internal reddening, star-formation 
spread (SFS), binary fraction, etc) may be somewhat subjective and unreliable, particularly 
when only photometry is available to work with. Examples of clusters evolving along
the early phase and characterised by such complex CMDs are abundant in the recent 
literature. For instance: NGC\,6611 (\citealt{N6611}), NGC\,4755 (\citealt{N4755}), 
NGC\,2244 (\citealt{N2244}), Bochum\,1 (\citealt{Bochum1}), Pismis\,5 and vdB\,80 
(\citealt{Pi5}), Collinder\,197 and vdB\,92 (\citealt{vdB92}).

More broadly, a robust determination of fundamental parameters of young 
clusters may provide important constraints for studies dealing with dynamical state 
and cluster dissolution time-scales (e.g. \citealt{GoodW09}; \citealt{Lamers10}), 
{\em infant mortality} (e.g. \citealt{LL2003}; \citealt{GoBa06}), star-formation 
rate (SFR) in the Galaxy (e.g. \citealt{LG06}; \citealt{SFR}), among others.

Over the years, several approaches have been proposed to tackle the important task
of finding reliable parameters of young clusters. Among these, \citet{NJ06} employ
a maximum-likelihood method to Hess diagram\footnote{Hess diagrams contain the 
relative density of occurrence of stars in different colour-magnitude cells of the 
Hertzsprung-Russell diagram (\citealt{Hess24}).} simulations (including binaries) to 
derive distances and ages. However, they do not consider differential reddening (DR),
and their method appears to be more efficient for clusters older than $\sim30$\,Myr.
\citet{Hill08} model CMDs of star forming regions and young star clusters by means 
of varying star-formation histories; because of confusion between signal and noise 
in CMDs, only marginal evidence for moderate age spreads was found. \citet{daRio2010}
add DR, age spreads and PMS stars to the \citet{NJ06} method, 
but adopt distance and reddening values from previous work. \citet{StHo2011} employ
Monte Carlo to estimate the age of ECs observed with near-infrared (UKIDSS) photometry. 
More recently, \citet{bruteF} (hereafter Paper\,I) included the smearing effect of 
photometric uncertainties on CMDs to approach the above issue by means of 
{\em brute-force}. However, because of computer limitations, the binary fraction, 
SFS, SFR, and the shape of the DR distribution had to be taken as fixed parameters.

In summary, the problem of obtaining reliable cluster parameters clearly lacks a 
comprehensive method that includes most (if not all) of the relevant conditions 
and parameters usually associated with the early cluster phases. Here we present 
a semi-analytical approach that 
follows along this direction. Instead of relying on {\em brute-force}, we now make
use of the global optimisation method known as adaptive simulated annealing (ASA, 
adapted from \citealt{ASA}) to search for the set of parameters that best reproduces
the photometric properties (i.e., the CMD or Hess diagram) of a cluster. The free 
parameters are the apparent distance modulus, foreground reddening, cluster mass, 
age, SFS, binary fraction, and DR, which 
are used to build the simulated Hess diagram for a given cluster. The function to 
be minimised by ASA is related to differences between the observed and simulated 
Hess diagrams. For conciseness, hereafter we will refer to our approach as \rr.

By explicitly including the binary fraction, SFS and the shape of the
DR distribution as free parameters on a semi-analytical approach, 
this work supersedes the procedure highlighted in Paper\,I.

\begin{figure}
\resizebox{\hsize}{!}{\includegraphics{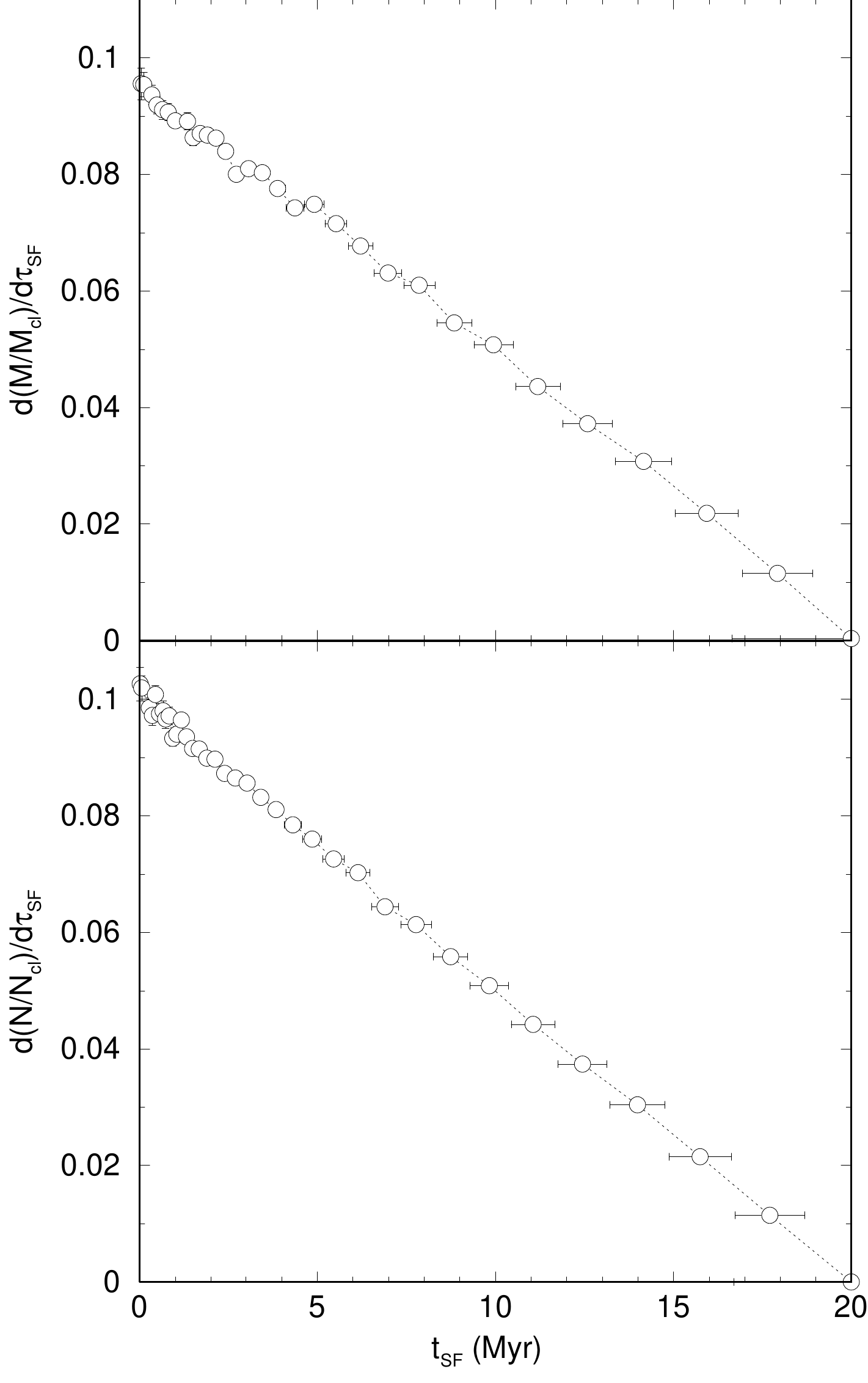}}
\caption[]{Top: The SFR (mass converted into stars with respect to the cluster mass) 
as a function of star-formation time for a cluster undergoing continuous star formation 
for 20\,Myr. Bottom: Same as above for the number of formed stars.}
\label{fig1}
\end{figure}

The present paper is organised as follows. In Sect.~\ref{ObsLim} we discuss the relevant 
effects that introduce observational limitations on CMDs. In Sect.~\ref{RecipeYC} we 
provide details on the cluster simulations. In Sect.~\ref{ASA} we briefly describe the 
\rr\ minimisation method, test it on model clusters and discuss the strategy adopted to 
obtain robust parameters. In Sect.~\ref{ProbAC} we apply \rr\ on actual young clusters, 
contextualise the results and discuss the low-mass cluster stochasticity. Concluding 
remarks are given in Sect.~\ref{Conclu}.

\section{Intrinsic limitations of CMDs}
\label{ObsLim}

Most of the problems related to obtaining reliable fundamental parameters of young clusters 
are illustrated in Fig.~\ref{fig1} of Paper\,I, in which a model CMD is subject to different 
values of distance modulus, DR and binary fraction. Clearly, when a moderate value of DR is 
added to the model CMD, the scatter and spread in the stellar sequences (resulting also from 
photometric uncertainties and binaries) tends to mask any intrinsic relationship between 
parameters of individual stars (e.g. the age \ts, mass \sms, magnitudes and colours) and 
those of the isochrones, an effect that increases significantly with distance modulus, DR 
amount and binary fraction. 

Following Paper\,I, we keep working with 2MASS\footnote{The Two Micron All Sky 
Survey, All Sky data release (\citealt{2mass06})} photometry and, in particular, with 
$\jj\times\jk$ CMDs (the colour least affected by photometric errors and the best 
discriminant for PMS stars - e.g. \citealt{Teu34}) and the corresponding Hess diagrams.
Reddening transformations are based on the absorption relations $A_J/A_V=0.276$ and 
$A_{K_S}/A_V=0.118$, with $R_V=3.1$ (\citealt{Cardelli89}; \citealt{DSB2002}). To avoid 
confusion between notations, we refer to the foreground reddening as \eJK, where
$\eJK=0.158\aV$, and to the differential reddening as \dV.

As shown in Paper\,I, the photometric uncertainties play an important role in shaping 
the CMD (or Hess diagram) morphology. Here we take them explicitly into account by 
assuming a normal (Gaussian) distribution of errors. Formally, if the magnitude (or colour) of a 
star is given by $\bar x\pm\sigma_x$, the probability of finding it at a specific value 
$x$ is given by $P(x)=\frac{1}{\sqrt{2\pi}\sigma_x}\,e^{{-\frac{1}{2}}\left(\frac{x-\bar x}
{\sigma_x}\right)^2}$. Thus, for each star we compute the fraction of the magnitude and 
colour that falls within a given cell of the Hess diagram (i.e., the cell density). 
Essentially, this corresponds to the difference of the error functions computed at the 
cell borders. By definition, the sum of the colour and magnitude density over all Hess 
cells should be the number of input stars. As a compromise between CMD resolution and 
computational time, the Hess diagrams used in this work consist of magnitude and colour 
cells of size $\Delta\jj=0.2$ and $\Delta\jk=0.02$, respectively. 

As a photometric quality criterion, we only keep stars with errors lower than 0.2, which 
also serves to emulate the photometric completeness function of the observations. The 
effect of this restriction on the luminosity function of a model cluster containing an 
arbitrarily large (for statistical purposes) number of stars can be seen in Fig.~\ref{fig1}
of Paper\,I.
After following the steps described below for assigning the mass, magnitudes, colours and 
uncertainties to each star, we apply the error restriction of keeping only stars with errors 
$\le 0.1$ and 0.2. Compared to the {\em complete} luminosity function, which presents a turnover 
at $\jj\approx16.9$, the restricted functions have turnovers that smoothly shift towards 
brighter magnitudes. 

\section{A simulation recipe for young clusters}
\label{RecipeYC}

Major changes with respect to the procedures adopted in Paper\,I for simulating young 
clusters are as follows: {\em (i)} The DR is described as a normal distribution, 
characterised by the mean \mdV\ and standard deviation around the mean \sdr, where 
both are free parameters. {\em (ii)} The SFR is a linearly decreasing function of 
time, which is a more realistic assumption, especially for clusters older than $\sim10$\,Myr 
(e.g. \citealt{Belloche11}; \citealt{Bate11} and references therein). 
Although somewhat arbitrary, this assumption has the advantage of requiring a single 
free parameter. In this case, it is just the cluster age (\ta), which we assume as the 
beginning of the star-forming event. For a cluster of mass \Mcl\ and age \ta, the 
SFR is the mass $M$ converted in stars as a function of time after the star formation 
start $t_{sf}$, which we express as 
$\frac{d\,M}{d\,t_{sf}} = \frac{2\,M_{clu}}{\ta}\left(1-t_{sf}/t_{clu}\right)$. This SFR 
is illustrated in Fig.~\ref{fig1}  (top panel), for a 20\,Myr old cluster that has been 
forming stars during the same 
time. Equivalently, the fraction of formed stars is also a linearly decreasing function 
of time (bottom). {\em (iii)} The SFS timescale (\sfs) is now a free parameter; 
instead of a fixed (and assumed to be equal to the cluster age) value, it can vary within
the range $0\le\sfs\le\ta$. Specifically, \sfs\ corresponds to the difference between
\ta\ and the minimum (isochrone) age compatible with the CMD stellar distribution. {\em (iv)} 
Binaries are expected to survive the early evolutionary 
phase of low-mass clusters, with the unresolved pairs being somewhat brighter and redder 
than the single stars, thus resulting in some broadening of the CMD stellar sequences, 
especially the MS (e.g. \citealt{NJ06}). We include them in our simulations by means of the 
(free) parameter \fB, which measures the fraction of unresolved binaries in a CMD. According 
to this definition, a CMD with $N_{CMD}$ detections and characterised by the binary fraction 
\fB, would have a number of individual stars expressed as $N_* = (1+\fB)N_{CMD}$. As an 
additional assumption, binaries are formed by pairing stars with the closest ages, regardless 
of the individual masses. This gives rise to a secondary to primary stellar mass-ratio 
($q=m_s/m_p$) that smoothly increases from very-low values up to $q\approx0.25$, and 
decreases for higher values of $q$ (Fig.~\ref{fig1} of Paper\,I).

%%%In addition, binaries also produce changes on the initial mass function of young, 
%massive star clusters (e.g. \citealt{Weidner09}). 

Thus, to simulate the CMD of a cluster on a given photometric system we: {\em (i)} Start 
with an artificial cluster of mass \Mcl, age \ta, SFS timescale \sfs, apparent distance 
modulus \mMJ, foreground reddening \eJK, mean DR \mdV\ and standard deviation \sdr, and 
binary fraction \fB. {\em (ii)} Assign each star an age according to $\ta-\sfs\le\ts\le\ta$, 
with a probability following a linearly decreasing age distribution (Fig.~\ref{fig1}). 
{\em (iii)} Select the individual stellar masses (\sms) by randomly taking values from 
\citet{Kroupa2001} mass function (with the upper mass value consistent with the age), 
until $\sum_i\sms_i=\Mcl$. {\em (iv)} Compute the magnitudes and colours of each star 
according to \ts\ and \sms, and the mass to light relations taken from the isochrone
corresponding to \ts. 
{\em (v)} Form the (unresolved) binary pairs according to \fB\ and compute their magnitudes 
and colours. {\em (vi)} For each star, randomly compute DR from a normal 
distribution characterised by the mean \mdV\ and standard deviation \sdr. {\em (vii)} Apply 
shifts in colour and magnitude according to the values of \eJK\ and \mMJ. {\em (viii)} Assign 
each artificial star a photometric uncertainty based on the average 2MASS errors and magnitude 
relationship. {\em (ix)} For more realistic representativeness, add some photometric noise to 
the stars. This step is taken to minimise the probability of stars with the same mass having 
exactly the same {\em observed} magnitude, colour, and uncertainty in the CMD. To do this, consider a star 
(of mass \sms\ and age \ts) with an intrinsic (i.e., measured from the corresponding isochrone)
magnitude $\overline{mag}$ and assigned uncertainty $\sigma_{mag}$. The noise-added magnitude $mag$ is 
then randomly computed from a normal distribution with a mean $\overline{mag}$ and standard deviation 
$\sigma_{mag}$. {\em (x)} Apply the same detection limit to the model CMD as for the observations, 
so that model and data share a similar photometric completeness function; in practise, this 
means that stars with photometric errors higher than 0.2 are discarded. {\em (xi)} Finally, 
build the corresponding $\jj\times\jk$ Hess diagram.

To minimise the critical stochasticity associated with low-mass clusters (see, e.g. 
Sect.~\ref{RnS}), steps {\em (ii)} to {\em (x)} are repeated for \nS\ (twin) clusters 
of mass \Mcl, age \ta, SFS timescale \sfs, and binary fraction \fB. Thus, each cell 
of the artificial Hess diagram (step {\em (xi)}) contains the density of stars 
averaged over the $N_{sim}$ clusters. Some technical details are described below.

The broken mass function of \citet{Kroupa2001} is defined as $dN/dm\propto m^{-(1+\chi)}$, 
with the slopes $\chi=0.3$ for $0.08\leq m(\ms)\leq0.5$ and $\chi=1.3$ for $m(\ms)>0.5$. 
The relations of photometric uncertainty with magnitude for the \jj\ and \ks\ 2MASS bands 
are well represented by $\sigma_J = 0.0214 + 2.48\times10^{-8}\exp{(J/1.071)}$ and 
$\sigma_{K_s}=0.0193+9.59\times10^{-9}\exp{(K_s/1.067)}$. These relations apply to the 
range $3\le\jj,\ks\le20$.

The stellar mass/luminosity relation is taken from the solar-metallicity isochrone sets of 
Padova\footnote{Built for the 2MASS filters at {\em http://stev.oapd.inaf.it/cgi-bin/cmd}.}
(\citealt{Girardi2002}) and \citet{Siess2000}. Both sets have been merged, since Padova 
isochrones should be used only for the MS (or more evolved sequences), and those of Siess 
apply to the PMS\footnote{As a caveat, some colour bias may be present in the Siess isochrones, 
because they use a single $T_e$-colour relation (from \citealt{KH95}) and do not take differences 
of the evolving surface gravities of PMS stars into account (private communication by the
referee M.G. Hoare, and M.S. Bessel).}. The merging occurs at the MS entry point, at $6.5\,\ms$ 
for the isochrones 
younger than 8\,Myr, $5.5\,\ms$ for 10\,Myr, $4.5\,\ms$ for 20\,Myr, and $3.5\,\ms$ for 
$\ge30$\,Myr. We now work with a rather high time resolution, considering isochrones with 
ages from 0 to 10\,Myr (with a step of 1\,Myr), 10 to 20\,Myr (step of 2\,Myr), and 20 to 
50\,Myr (step of 5\,Myr). The cluster models consist of stars more massive than $0.1\,\ms$, 
the lowest available mass in the PMS isochrones. Magnitudes, colours and mass for model stars 
with intermediate age values are obtained by interpolation among the neighbouring isochrones. 
Also, the maximum stellar mass present in the adopted isochrones ranges from 60\,\ms\ 
(at 0.2\,Myr) through 36\,\ms\ (5\,Myr), 19\,\ms\ (10\,Myr), 9\,\ms\ (30\,Myr), and 7.3\,\ms\ 
(50\,Myr). 

%\footnote{The updated PMS isochrones
%present some differences with respect to those used in \citet{bruteF}, especially at
%the very-low mass regime. The differences are expected to affect mainly the distance
%modulus.}

\section{Parameter optimisation with \rr}
\label{ASA}

As discussed in the previous section, the problem now is to search for the set of 8 
free parameters $\P\equiv\left\{\Mcl, \ta, \sfs, \mMJ, \eJK, \mdV, \sdr,\fB\right\}$
that produces the best match between the observed and simulated Hess diagrams of a 
given star cluster. This is achieved by minimising the root mean squared residual
between both diagrams (\x2) that, by construction, is a function of the free parameters,
i.e., $\x2=\x2(\P)$. As shown in Paper\,I, such a task would take an extremely long time 
for the {\em brute-force} method, especially when one considers a realistic (relatively 
wide) range of values for such an extended set of parameters. 

Among the several optimisation methods available in the literature, the adaptive simulated 
annealing (ASA) appears to suit our purposes, because it is relatively time efficient and 
robust (e.g. \citealt{ASA}). ASA has been shown to be a global optimisation technique that 
distinguishes between different local minima, and the \x2\ hyper-surface is characterised 
by the presence of such features (Paper\,I). Simulated annealing derives its name from 
the metallurgical process by which the controlled heating and cooling of a material is used 
to increase the size of its crystals and reduce their defects. If an atom is stuck to a local
minimum of the internal energy, heating forces it to randomly wander through higher-energy 
states. In the present context, a state is a simulation corresponding to a specific set of 
values of the parameters to be optimised. Then, the slow cooling increases the probability 
of finding states of lower energy than the initial one.

Minimisation starts by the definition of individual search-ranges (wide enough to permit 
the occurrence of any value compatible with the cluster nature) and variation steps for 
each free parameter; next, an initial point is randomly selected and the starting residual 
value $R^i_{rms}$ is computed. At this point, we define the initial {\em temperature} as
$\mathcal{T}\equiv R^i_{rms}$. Then ASA takes a step (i.e, changing the initial parameters) 
and the new value $R^{i+1}_{rms}$ is evaluated. Specifically, this implies that a new Hess 
diagram has to be simulated with the new parameters. By definition, any downhill ($R^{i+1}_{rms}
<R^i_{rms}$) step is accepted, with the process repeating from this new point. However, uphill 
moves may also be taken, with the decision made by the Metropolis (\citealt{Metropolis53}) 
criterion, which allows small uphill moves while rejecting large ones, thus enabling ASA 
to escape from local minima. When an uphill move is required, ASA computes the value of 
$\mathcal{P} = e^{-(\Delta/\mathcal{T})}$, where $\Delta\equiv R^{i+1}_{rms}-R^i_{rms}$. 
$\Delta/\mathcal{T}$ is positive in an uphill move and so, $\mathcal{P}$ is a number between 
0 and 1 that is compared with a random number $0\le\mathcal{N}\le1$. If $\mathcal{P}\ge\mathcal{N}$, 
the uphill move is accepted and the algorithm moves on from that point; in case of rejection, 
another point is chosen for a trial evaluation of $\mathcal{P}$ and $\mathcal{N}$. Clearly, 
large values of $\Delta$ and low $\mathcal{T}$ make acceptance of an uphill move less likely. 
After each successful move, the {\em temperature} of the system is reduced according to 
$\mathcal{T}\rightarrow f\,\mathcal{T}$, which makes uphill moves less likely to be accepted 
as ASA focuses upon the most promising area for optimisation. For a more precise (but slower) 
convergence rate, we use $f=0.95$. Variation steps decrease as the minimisation is successful 
and ASA closes in on the global minimum ($\P\rightarrow\Pm$). The termination criterion 
for a run occurs when $\Delta\le10^{-6}$ or, for runtime sake, the number of trial function 
evaluations - for a single uphill move - reaches $2.5\times10^5$.

We define the root mean squared residual between observed ($H_{obs}$) and simulated 
($H_{sim}$) Hess diagrams, composed of $n_c$ and $n_m$ colour and magnitude cells, as

\begin{equation}
\label{eq1}
\x2=\sqrt{\frac{1}{N_{obs}}\sum_{i,j=1}^{n_c,n_m}\frac{\left[H_{obs}(i,j)-H_{sim}(i,j)\right]^2}
{\left[H_{obs}(i,j)+H_{sim}(i,j)\right]}}~~~.
\end{equation}

The sum is restricted to non-empty cells, and the normalisation by the simulated$+$observed 
density of stars in each cell gives a higher weight to more populated cells\footnote{In Poisson 
statistics, where the uncertainty of a measurement $\mathcal{N}$ is $\sigma=\sqrt{\mathcal{N}}$, 
our definition of \x2\ turns out being somewhat equivalent to the usual $\chi^2$.}. Finally, 
the squared sum is divided by the total number of stars in the observed CMD, which makes \x2\ 
dimensionless, and preserves the number statistics when comparing clusters with unequal numbers 
of stars. 

The adopted strategy is: we start by building the Hess diagram corresponding to the 
$\jj\times\jk$ CMD of an actual young cluster. Next, we minimise Eq.~\ref{eq1} to find 
the values of \Pm\ that give rise to the simulated Hess diagram that best matches the 
observation. Before reaching the absolute minimum, several ASA iterations are undertaken, 
which generates new starting points. At each new point (i.e, a new set of values \P) 
produced by ASA, steps {\em (i)} to {\em (xi)} of Sect.~\ref{RecipeYC} have to be repeated.

By construction, the output of a single ASA run is the set of optimum values \Pm\ (expected to
correspond to the global minimum) and the respective \x2, but with no reference to uncertainties. 
However, given the statistical nature of the process used to build the simulated Hess diagrams 
- as well as the stochasticity intrinsically associated to young clusters, uncertainties 
to the parameters are strongly required. Thus, we run ASA several times (\nR) and compute 
the weighted mean of the parameters over all runs, using the \x2\ of each run as weight
($w=1/\x2$). It is important to remark that each new run begins with a totally new point 
\P, with values randomly selected within the individual search-ranges (see above); this strategy 
provides a means to minimise biases related to fixed initial conditions.
Besides uncertainties, this procedure also provides a means to check for convergence 
patterns (see below). In summary, the \rr\ approach consists of the full set of procedures 
so far described, namely, the star cluster simulation (Sect.~\ref{RecipeYC}), the \nR\ ASA 
runs plus the statistical analysis (see above). Ideally, the parameters produced by \rr\ 
should present a lower dispersion around the mean as \x2\ declines.

\subsection{Tests with model star clusters}
\label{ConExp}

\begin{figure}
\resizebox{\hsize}{!}{\includegraphics{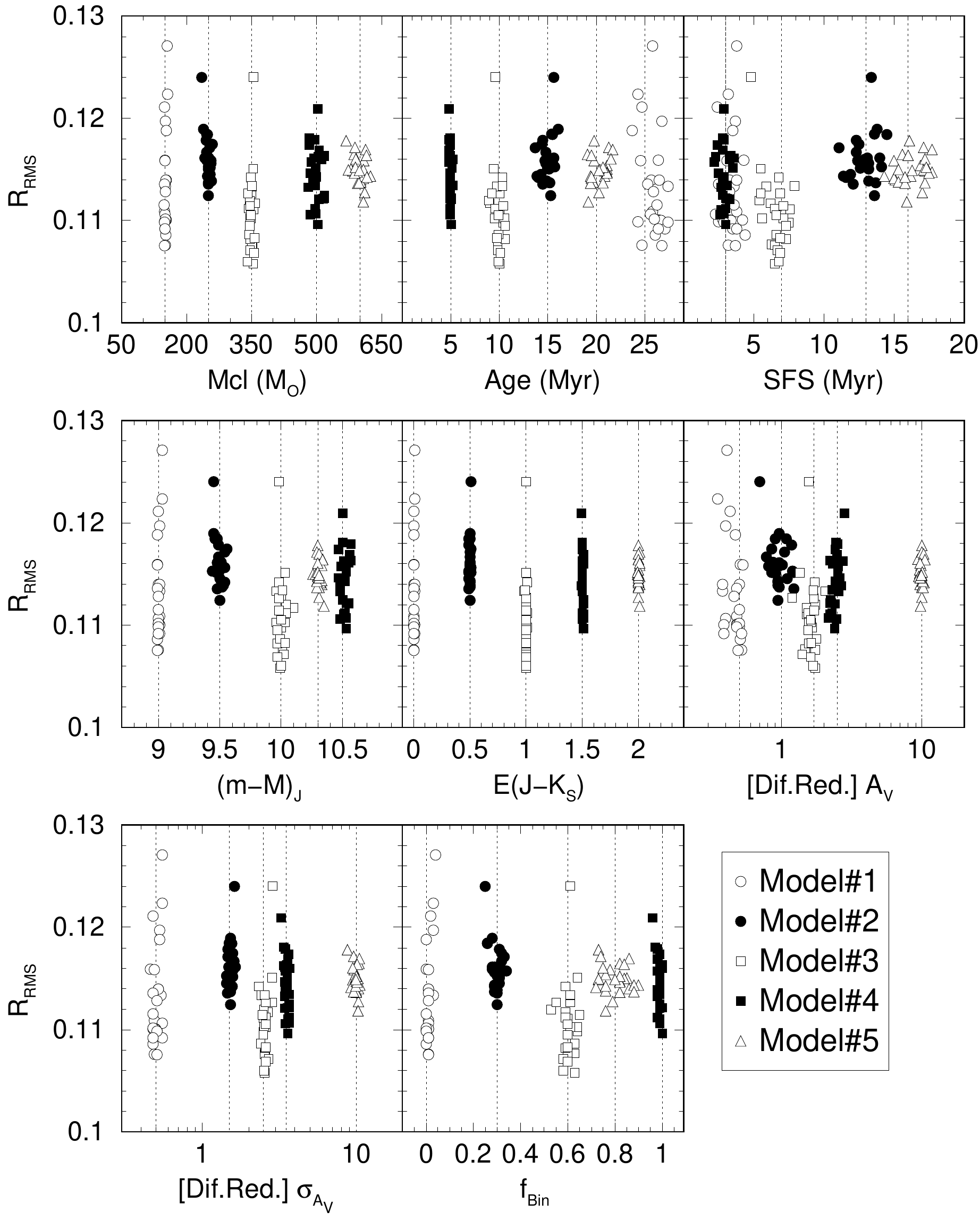}}
\caption[]{Convergence pattern for the models in Table~\ref{tab1}, with the
input values shown by the vertical lines. Each symbol represents a single run.}
\label{fig2}
\end{figure}

We now employ template CMDs built with typical parameters of  PMS-rich young clusters 
to test the ability of \rr\ in recovering the input values. As 
described in Table~\ref{tab1}, the models cover a broad range of properties, with 
$150\le\Mcl(\ms)\le600$, $5\le\ta (Myr)\le25$, $3\le\sfs(Myr)\le16$, $0\le\fB\le1$,
including high DR absorption values (Model\#5). In all 
cases we consider $\nS=100$ and $\nR=25$ ASA runs. The retrieved parameters, together 
with the corresponding \x2, are listed in Table~\ref{tab1}. For simplicity, Table~\ref{tab1}
lists only the parameters found for the runs corresponding to the minimum, maximum and mean 
values of \x2. Note that the mean values are obtained by averaging out the \nR\ independent
outputs and using the respective \x2\ as weight; usually, the standard deviations are lower
than 5\% with respect to the mean values. The complete set of solutions is shown in 
Fig.~\ref{fig2}. 

\begin{figure}
\resizebox{\hsize}{!}{\includegraphics{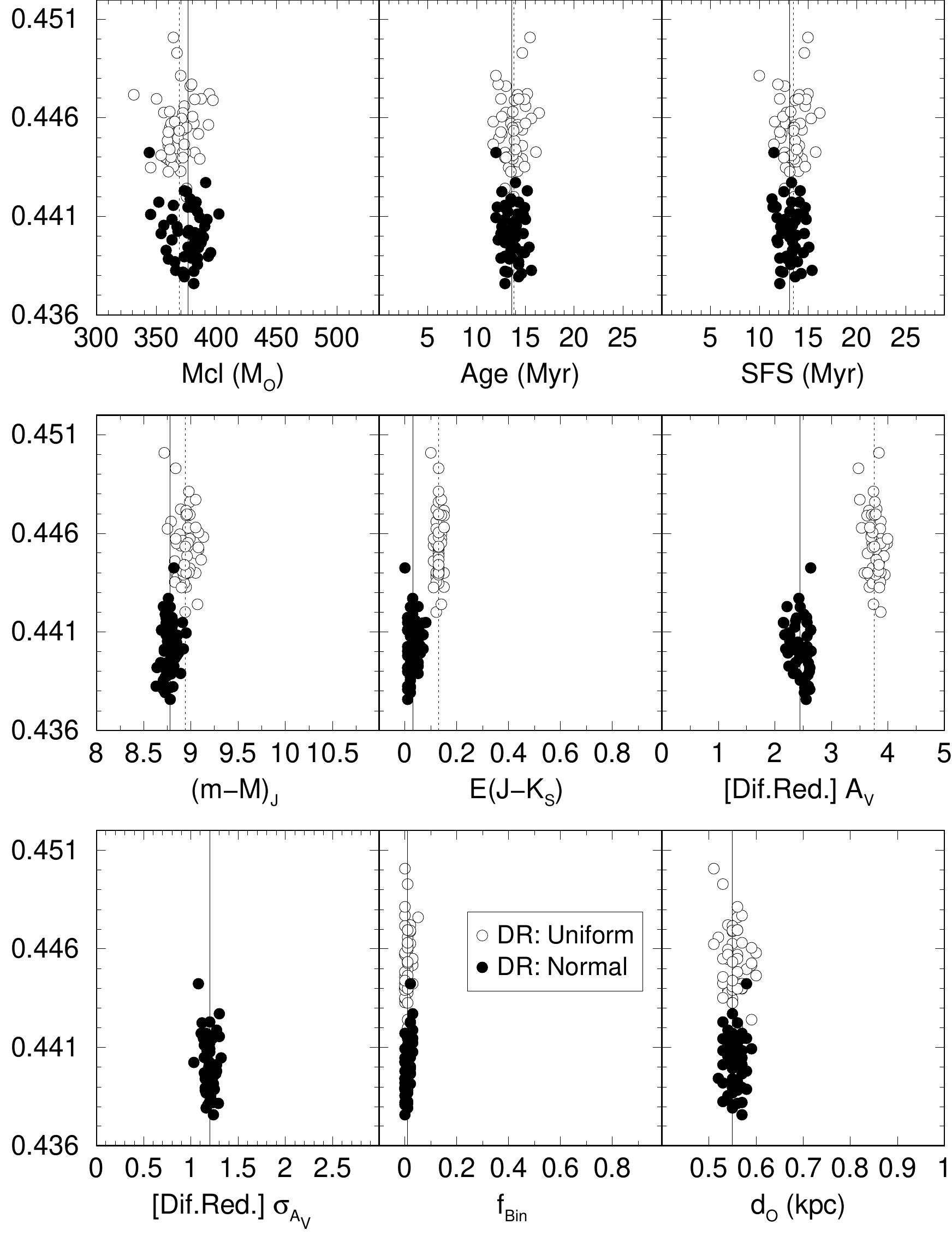}}
\caption[]{Similar to Fig.~\ref{fig2} for Collinder\,197. Results for the uniform
(empty symbols) and normal (filled) differential reddening modes are shown. Vertical
lines correspond to the mean values for the uniform (dotted) and normal (solid) 
distributions. The abscissa scales represent the search-ranges used at the beginning
of the ASA minimisation process.}
\label{fig3}
\end{figure}

\begin{table*}
\caption[]{Recovery of model parameters with \rr}
\label{tab1}
%\tiny
\renewcommand{\tabcolsep}{1.7mm}
\renewcommand{\arraystretch}{1.25}
\begin{tabular}{ccccccccccc}
\hline\hline
Rank&$\x2$&$M_{clu}$& Age & \sfs &\mMJ & \eJK&\mdV  &\sdr&\fB&$M_{CMD}$\\
    &     &  (\ms)  &(Myr)&(Myr)&(mag)&(mag)&(mag)&  (mag) &   &(\ms)    \\
(1)&(2)&(3)&(4)&(5)&(6)&(7)&(8)&(9)&(10)&(11)\\
\hline
\multicolumn{2}{c}{Model\#1}& 150 & 25 & 3 & 9.0 & 0.0 & 0.5 & 0.5 & 0.0 &---\\
\hline
Min &0.1556& 149 & 26.8 &  3.7 & 8.99 & 0.00 & 0.49 & 0.51 & 0.01 & 133 \\
Max &0.1763& 155 & 25.8 &  3.8 & 9.03 & 0.01 & 0.41 & 0.55 & 0.04 & 145 \\
Mean&0.1653& $151\pm2$ & $25.9\pm1.0$ & $3.3\pm0.6$ & $9.00\pm0.01$ & 
          $0.01\pm0.01$ & $0.46\pm0.05$ & $0.51\pm0.03$ & $0.01\pm0.01$ & $135\pm3$ \\
          
\hline
\multicolumn{2}{c}{Model\#2}& 250 & 15 & 13 & 9.5 & 0.5 & 1.0 & 1.5 & 0.3 &---\\
\hline
Min &0.1124& 250 & 15.3 & 13.6 & 9.50 & 0.50 & 0.94 & 1.53 & 0.30 & 153 \\
Max &0.1240& 235 & 15.6 & 13.4 & 9.45 & 0.51 & 0.70 & 1.62 & 0.25 & 142 \\
Mean&0.1160& $250\pm6$ & $15.0\pm0.6$ & $12.9\pm0.7$ & $9.50\pm0.03$ & 
          $0.50\pm0.01$ & $0.97\pm0.13$ & $1.53\pm0.05$ & $0.30\pm0.02$ & $153\pm5$ \\

\hline
\multicolumn{2}{c}{Model\#3}& 350 & 10 & 7 & 10.0 & 1.0 & 1.7 & 2.5 & 0.6 &---\\
\hline
Min &0.1058& 354 & 10.0 & 6.5 & 9.99 & 1.00 & 1.73 & 2.54 & 0.63 & 200 \\
Max &0.1240& 355 &  9.6 & 4.8 & 9.98 & 1.00 & 1.57 & 2.87 & 0.61 & 204 \\
Mean&0.1105& $349\pm5$ & $9.9\pm0.5$ & $6.6\pm0.7$ & $10.00\pm0.03$ & 
          $1.00\pm0.01$ & $1.61\pm0.16$ & $2.57\pm0.12$ & $0.60\pm0.03$ & $193\pm3$ \\

\hline
\multicolumn{2}{c}{Model\#4}& 500 & 5 & 3 & 10.5 & 1.5 & 2.5 & 3.5 & 1.0 &---\\
\hline
Min &0.1096& 503 & 5.0 & 3.0 & 10.53 & 1.51 & 2.39 & 3.59 & 1.00 & 278 \\
Max &0.1209& 503 & 4.8 & 2.9 & 10.50 & 1.49 & 2.81 & 3.25 & 0.96 & 261 \\
Mean&0.1143& $500\pm11$ & $4.9\pm0.1$ & $2.9\pm0.3$ & $10.52\pm0.03$ & 
          $1.50\pm0.01$ & $2.40\pm0.17$ & $3.53\pm0.11$ & $0.99\pm0.01$ & $280\pm6$ \\

\hline
\multicolumn{2}{c}{Model\#5}& 600 & 20 & 16 & 10.3 & 2.0 & 10.0 & 10.0 & 0.8 &---\\
\hline
Min &0.1119& 609 & 19.2 & 15.9 & 10.35 & 2.01 & 9.77 & 10.28 & 0.76 & 154 \\
Max &0.1178& 569 & 19.8 & 16.1 & 10.30 & 2.00 & 9.97 &  8.80 & 0.73 & 140 \\
Mean&0.1149& $599\pm14$ & $20.3\pm0.8$ & $16.3\pm1.0$ & $10.31\pm0.02$ & 
          $2.00\pm0.01$ & $9.93\pm0.26$ & $9.95\pm0.40$ & $0.80\pm0.07$ & $152\pm4$ \\
\hline
\end{tabular}
\begin{list}{Table Notes.}
\item Cols.~(1) and (2): \x2\ rank and value; Col.~(3): actual cluster mass; Col.~(4):
cluster age; Col.~(5): SFS timescale; Col.~(6): apparent distance modulus in the \jj\ 
band; Col.~(7): foreground reddening; Cols.~(8) and (9): mean DR and standard deviation; 
Col.~(10): unresolved binary fraction; Col.~(11): cluster mass as measured in the CMD. 
The average stellar mass of the models is $\overline{m_*}\approx0.6\,\ms$.
\end{list}
\end{table*}

As expected, the retrieved parameters close in on the input values as \x2\ declines, 
although with somewhat different dispersions around the mean among the parameters. The 
convergence pattern is clearly tighter for \Mcl, \eJK, and \mMJ; followed by \sdr\ and 
\ta, with \fB, \sfs, and \mdV\ at a third level. This is consistent with the mean 
values and respective standard deviations listed in Table~\ref{tab1}. Nevertheless, it 
is remarkable that \rr\ can retrieve the input values even for a CMD with such a high 
DR amount as that in Model\#5. Also, its ability to disentangle DR and binary fraction 
comes from the fact that, although both effects tend to redden the stellar sequences, 
binaries brighten them, while DR shifts them towards the opposite direction. It is 
also interesting the strong sensitivity for very short (e.g. Models\#1 and 4) and long 
(Models\#2 and 5) SFSs. 

Similarly to Paper\,I, we also compute the cluster mass that potentially would be measured 
based on CMD properties ($M_{CMD}$), which is an important piece of information for, e.g., 
establishing the dynamical state of a cluster. The first step is to find the actual mass 
of each star occurring in the CMD. However, this may be very difficult to find, especially 
in the presence of DR, SFS, unresolved binaries, and photometric 
uncertainties (which decreases the number of stars that remain detectable in a CMD as 
the distance modulus increases). In practise, what is usually done is: having derived 
the values of age, distance modulus and foreground reddening, $M_{CMD}$ is computed by 
finding the probable mass for each star in the CMD by interpolation of the observed 
colour and magnitude among those of the nearest isochrones. Obviously, 
the precision of this procedure relies heavily on the amount of DR, 
photometric noise, binaries, etc. For instance, a heavily differentially reddened cluster
of $\Mcl\sim600\,\ms$, characterised by a moderate binary fraction and distance modulus, 
would have only $\approx25\%$ of its actual mass estimated based on CMD properties (e.g.,
Model\#5 in Table~\ref{tab1}). 

We close this section by concluding that \rr\ - the minimisation of residuals between 
the observed and simulated Hess diagrams by means of ASA - is efficient in retrieving 
the input parameters of model CMDs that cover a variety of conditions.

\section{Probing actual star clusters}
\label{ProbAC}

Based on the experience gained in Sect.~\ref{ConExp} with model CMDs, we now move 
on to investigate actual young clusters with \rr. For this we use Collinder\,197 
(\citealt{vdB92}) and Pismis\,5 (\citealt{Pi5}). Both clusters have been studied 
in Paper\,I, which thus will allow us to compare the effect of different assumptions 
on SFR, DR and binaries, on the derived parameters. They are projected near the 
Galactic equator ($b\approx1\degr$) and have CMDs dominated by faint stars. 
To minimise confusion between intrinsic PMS stars and red dwarfs of the Galactic 
field, we build field-star decontaminated CMDs by means of the algorithm developed 
in \citet{BB07} and improved in \citet{vdB92}. As a result, the decontaminated CMD 
of Collinder\,197 has 690 stars (essentially PMS), while Pismis\,5 has only 101, which 
is important for examining the effect of CMDs with different numbers of stars on our 
analysis. In addition, the colour-colour diagrams of both objects (Collinder\,197:
Fig.~7 of \citealt{vdB92}; Pismis\,5: Fig.~9 of 7\citealt{Pi5}) do not contain
detections with abnormally high infrared excesses that might be due to circumstellar 
material.

\begin{figure}
\resizebox{\hsize}{!}{\includegraphics{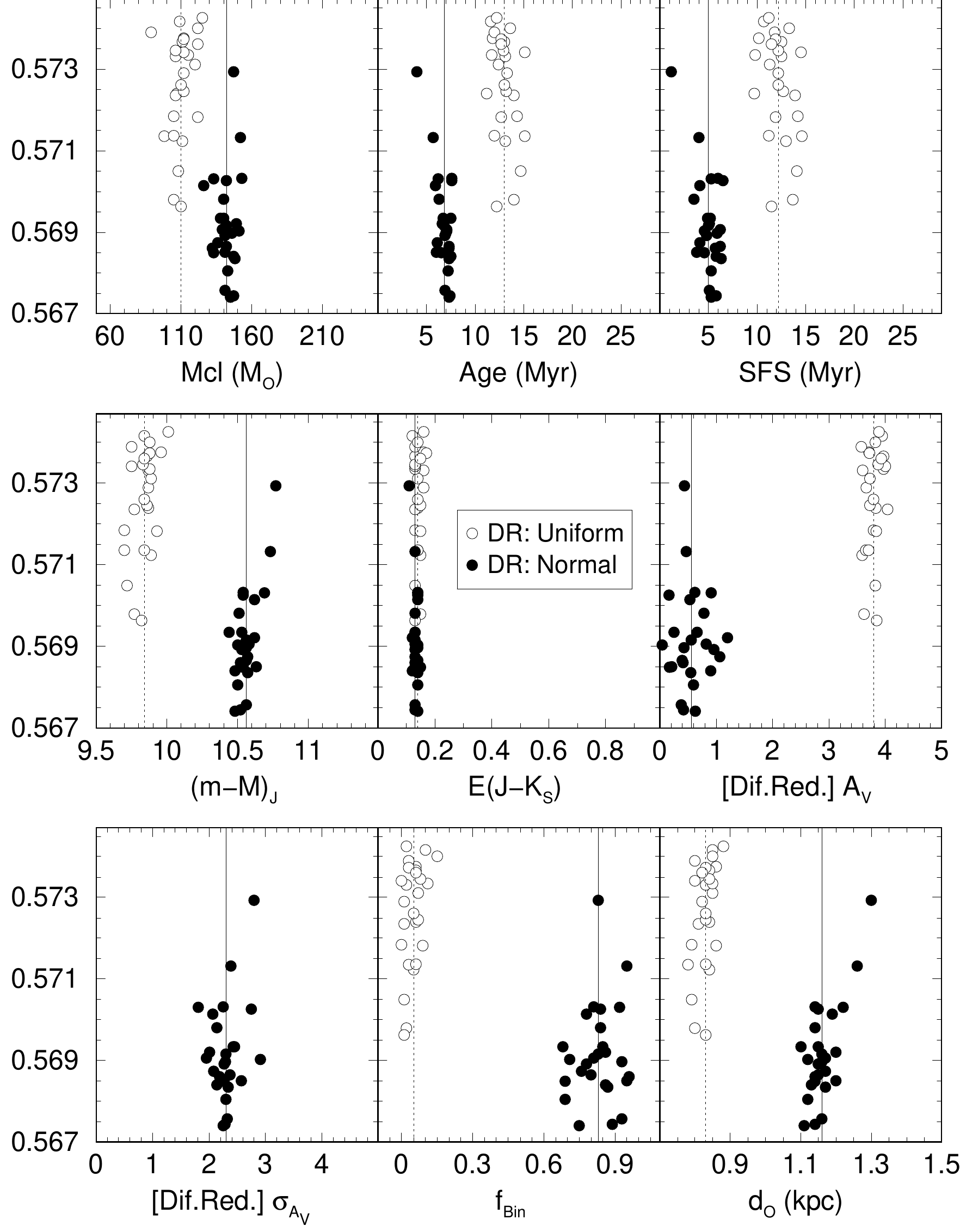}}
\caption[]{Same as Fig.~\ref{fig3} for Pismis\,5. }
\label{fig4}
\end{figure}

After several tests with \rr, we settled on $\nS=50$ and $\nR=50$ for Collinder\,197 
and $\nS=500$ and $\nR=25$ for Pismis\,5 as a compromise between convergence pattern 
and running time\footnote{As a technical note we remark that each ASA run for Collinder\,197 
took $\sim1.2$\,hours, and $\sim2.5$\, hours for Pismis\,5, on a single core of an {\em 
Intel Core i7 920@2.67\,GHz} processor.}. Besides the normal DR distribution, for comparison 
purposes we also consider the case of a uniform (or flat) distribution. The parameters 
obtained with \rr\ are given in Table~\ref{tab2}, and the full set of runs is shown in 
Figs.~\ref{fig3} and \ref{fig4}. For comparison with other clusters, we also compute the 
bolometric magnitude and mass to light ratio. We remark that it is significant that convergence
patterns reached after several thousand ASA iterations, similar to those of the model 
CMDs, occur for all parameters of both clusters.

\begin{table*}
\caption[]{Parameters of Collinder\,197 and Pismis\,5}
\label{tab2}
\tiny
\renewcommand{\tabcolsep}{0.6mm}
\renewcommand{\arraystretch}{1.5}
\begin{tabular}{ccccccccccccc}
\hline\hline
$\x2$&$M_{clu}$& Age & \sfs &\mMJ & \eJK&\mdV  &\sdr &\fB&$M_{CMD}$&\ds  
    &$10^3$\,MLR& $M_{BOL}$\\
    &  (\ms)  &(Myr)&(Myr)&(mag)&(mag)&(mag)&(mag)&   &  (\ms)  &(kpc)
    &($\ms/L_\odot$)& (mag) \\
(1)&(2)&(3)&(4)&(5)&(6)&(7)&(8)&(9)&(10)&(11)&(12)&(13)\\

\hline
&\multicolumn{11}{c}{Collinder\,197 - Differential reddening mode: normal}\\
\hline
0.4376& 381 & 12.9 & 12.1 & 8.78 & 0.01 & 2.55 & 1.24 & 0.00 & 211 & 0.57 & 4.7 & -7.5\\
0.4442& 344 & 12.0 & 11.5 & 8.82 & 0.00 & 2.63 & 1.08 & 0.02 & 182 & 0.58 & 6.6 & -7.0 \\
0.4401& $376\pm13$ & $13.6\pm0.9$ & $13.1\pm1.0$ & $8.78\pm0.07$ & $0.03\pm0.02$ & $2.44\pm0.15$ &
          $1.20\pm0.06$ & $0.01\pm0.01$ & $208\pm7$ & $0.55\pm0.02$ & $4.2\pm1.0$ & $-7.6\pm0.2$\\

$\Delta1\sigma$& $376^{+230}_{-190}$ & $12.9^{+7.2}_{-4.7}$ & $12.1^{+0.7}_{-5.3}$ & $8.78^{+1.30}_{-0.76}$ 
   & $0.01^{+0.17}_{-0.01}$ & $2.55^{+1.02}_{-0.96}$ & $1.24^{+0.96}_{-0.51}$ & $0.00^{+0.39}_{-0.00}$ 
   & $211^{+30}_{-56}$ & $0.57^{+0.33}_{-0.17}$ & $4.7^{+3.3}_{-0.4}$ & $-7.5^{-0.1}_{+0.6}$\\

$\overline{1\sigma}$& $396\pm96$ & $14.3\pm3.5$ & $10.0\pm1.7$ & $8.95\pm0.47$ & $0.01\pm0.01$ & $2.58\pm0.50$ &
          $1.42\pm0.40$ & $0.17\pm0.11$ & $224\pm12$ & $0.58\pm0.12$ & $3.5\pm1.9$ & $-7.9\pm0.6$\\
          
\hline
&\multicolumn{11}{c}{Collinder\,197 - Differential reddening mode: uniform}\\
\hline
0.4420& 375 & 13.8 & 13.6 & 8.94 & 0.12 & 3.87 & --- & 0.01 & 225 & 0.56 & 5.3 & -7.4\\
0.4501& 364 & 15.5 & 15.0 & 8.72 & 0.10 & 3.84 & --- & 0.00 & 224 & 0.51 & 4.7 & -7.5 \\
0.4453& $369\pm12$ & $13.8\pm1.1$ & $13.5\pm1.1$ & $8.94\pm0.09$ & $0.13\pm0.01$ & 
     $3.76\pm0.11$ & --- & $0.01\pm0.01$ & $227\pm8$ & $0.55\pm0.02$ & $4.1\pm0.8$ & $-7.6\pm0.2$\\          

\hline
&\multicolumn{11}{c}{Pismis\,5 - Differential reddening mode: normal}\\
\hline
0.5674& 145 & 7.3 & 5.3 & 10.48 & 0.14 & 0.63 & 2.25 & 0.75 & 80 & 1.11 & 3.1 & -6.9\\
0.5729& 147 & 4.0 & 1.2 & 10.77 & 0.11 & 0.44 & 2.80 & 0.83 &100 & 1.30 & 1.8 & -7.5 \\
0.5692& $142\pm7$ & $6.8\pm0.8$ & $5.0\pm1.0$ & $10.56\pm0.08$ & 
          $0.13\pm0.01$ & $0.56\pm0.29$ & $2.30\pm0.25$ & $0.83\pm0.08$ & $81\pm4$ & 
          $1.16\pm0.04$ & $2.6\pm0.4$ & $-7.1\pm0.2$\\

$\Delta1\sigma$& $145^{+91}_{-83}$ & $7.3^{+8.0}_{-3.8}$ & $5.3^{+1.8}_{-2.7}$ & $10.48^{+0.82}_{-0.72}$ 
   & $0.14^{+0.06}_{-0.07}$ & $0.63^{+1.25}_{-0.63}$ & $2.25^{+1.66}_{-0.85}$ & $0.75^{+0.25}_{-0.57}$ 
   & $80^{+6}_{-23}$ & $1.11^{+0.44}_{-0.25}$ & $3.1^{+2.2}_{-0.4}$ & $-6.9^{-0.1}_{+0.6}$\\

$\overline{1\sigma}$& $154\pm50$ & $9.0\pm3.2$ & $5.0\pm1.3$ & $10.52\pm0.38$ & 
          $0.14\pm0.04$ & $0.92\pm0.54$ & $2.62\pm0.70$ & $0.73\pm0.16$ & $82\pm7$ & 
          $1.12\pm0.20$ & $2.7\pm1.5$ & $-7.0\pm0.6$\\
          
\hline
&\multicolumn{12}{c}{Pismis\,5 - Differential reddening mode: uniform}\\
\hline
0.5696& 110 & 12.2 & 11.5 &  9.82 & 0.13 & 3.85 & ---  & 0.01 & 57 & 0.83 & 4.5 & -6.2\\
0.5743& 125 & 12.2 & 11.2 & 10.01 & 0.16 & 3.89 & ---  & 0.02 & 63 & 0.88 & 4.5 & -6.4 \\
0.5726& $110\pm8$ & $13.0\pm1.1$ & $12.2\pm1.2$ & $9.84\pm0.08$ & 
          $0.14\pm0.01$ & $3.79\pm0.13$ & --- & $0.05\pm0.04$ & $57\pm4$ & $0.83\pm0.03$ & 
          $4.4\pm0.9$ & $-6.3\pm0.2$\\          
          
\hline
\end{tabular}
\begin{list}{Table Notes.}
\item Col.~(1): minimum, maximum and mean \x2\ values; Col.~(2): actual cluster mass; 
Col.~(3): cluster age; Col.~(4): SFS; Col.~(5): apparent distance 
modulus in the \jj\ band; Col.~(6): foreground reddening; Cols.~(7) and (8): differential 
reddening; Col.~(9): unresolved binary fraction; Col.~(10): cluster mass as measured in 
the CMD; Col.~(11): distance from the Sun; Col.~(12): bolometric mass to light ratio;
Col.~(13): bolometric magnitude. $\Delta1\sigma$: parameters occurring at the boundaries of the 
$1\sigma$ domain. $\overline{1\sigma}$: weighted average of values occurring within the 
$1\sigma$ domain.
\end{list}
\end{table*}

In both cases, the \x2\ corresponding to the normal DR distribution tends to be lower than 
those of the uniform distribution. Also, reflecting the larger number of CMD stars, the \x2\ 
values of Collinder\,197 are significantly lower than those of Pismis\,5. 

Before moving on to interpreting the results, it is important to remind that each \x2\ value 
and corresponding optimum parameters result from several thousand iterations, as ASA searches 
the \x2\ hyper-surface for the absolute minimum of Eq.~\ref{eq1}. At each iteration, \nS\ twin 
clusters (i.e., consisting of exactly the same set of parameters) are built and incorporated 
into the simulated Hess diagram. Thus, the occurrence of a tight convergence pattern for 
the absolute minimum parameters over a series of independent runs shows a self consistency of 
the method and cannot be taken as fortuitous or 
model dependent. Instead, it would be strongly indicative that the optimum parameters produced 
by \rr\ are indeed representative of those of the cluster being studied. On the other hand, we
remark that this argument applies only to the absolute minimum of each \rr\ run, since the \x2\ 
topology around this feature is not taken into account. In this sense, the quoted errors should 
be taken as internal, probably not reflecting the realistic parameter uncertainties. Indeed, as 
we show in Sect.~\ref{Min}, in some cases the drop towards the absolute minimum is somewhat 
gentle, which means that there can be a significant dispersion (different values at similar 
\x2\ levels) around the absolute minima. Although the depression shape departs from gaussianity, 
we estimate the approximate extension of the $1\sigma$ domain and compute the weighted mean and 
dispersion for the parameters occurring inside it (again using the individual \x2\ as weight). 
The results - restricted to the uniform DR - are given in the additional entry labelled as 
$\overline{1\sigma}$ in Table~\ref{tab2}. Compared to the previous statistics, the uncertainties 
now are more realistic and consistent both with the \x2\ topology and the assumptions incorporated
into the simulations. For completeness, we also provide in Table~\ref{tab2} the parameter values 
that occur at the boundaries of the $1\sigma$ domain (labelled as $\Delta1\sigma$). In general,
they imply considerably larger errors than before. However, given the underlying assumption of
residual minimisation and respective \x2\ value of different parameter sets, we believe that the 
weighted-mean values ($\overline{1\sigma}$) are more representative for the clusters dealt with
in this work. In addition, the mean values are essentially unchanged, except for a somewhat higher 
binary fraction for Collinder\,197.

\paragraph{\tt Collinder\,197:} The values of cluster mass, age, SFS, binary fraction, and 
distance from the Sun are essentially insensitive to the DR mode. They consistently indicate 
a cluster of $\Mcl\approx400\,\ms$, $\ta\approx14$\,Myr, $\sfs\approx10$\,Myr, with a relatively 
low binary fraction of $\fB\la0.2$, located at $\approx0.6$\,kpc from the Sun. However, assuming 
that representativeness increases for lower \x2\ values, DR in this cluster would follow a normal 
distribution characterised by $\mdV\approx2.6$ and $\sdr\approx1.4$. The differences that occur in 
distance modulus and foreground reddening can be accounted for by the single, fixed value of \dV\ 
of the uniform DR distribution, which requires higher values of \mMJ\ and \eJK\ than the normal 
mode to account for the observed colour spread of the stellar sequences. The foreground reddening 
is very low, corresponding to $\aV\approx0.1$.

\paragraph{\tt Pismis\,5:} Compared to Collinder\,197, here the DR modes result in significant 
differences for most of the cluster parameters, except for the foreground reddening. Also, as 
expected from the lower number of stars, the convergence patterns are in general looser than 
for Collinder\,197. Perhaps the most contrasting result lies in the binary fraction that reaches
$\fB\approx0.7$. Lacking  one degree of freedom, the uniform DR mode tries to describe the CMD 
spread by assuming a large value of \mdV\ with essentially no binaries. In contrast, the normal 
mode requires a lower \mdV\ with a high dispersion \sdr, which yields a high binary fraction. 
Combined, the low \mdV\ and high \fB\ of the normal mode put Pismis\,5 at a distance $\sim40\%$ 
larger than that implied by the uniform mode. The foreground reddening, $\aV\approx0.9$, is higher 
than that of Collinder\,197.

Interestingly, because of the additional free parameter, the normal DR distribution 
produces a cluster age significantly younger than that implied by the uniform distribution, 
especially for Pismis\,5. Also, in both cases the SFS is equivalent to $\sim60\%$ of the cluster 
age.

Comparing the values obtained by \rr\ with those in Paper\,I, we see that now the ages 
tend to be somewhat older and the distances shorter, with a reasonable agreement among 
the other parameters. A similar conclusion applies to the scarce works on both 
objects. Previous estimates for Collinder\,197 are: age$\sim5\pm4$\,Myr, 
$\ds\sim1.1\pm0.2$\,kpc, and mass $100-500\,\ms$ (\citealt{vdB92}, and references therein).
For Pismis\,5, they are: age $5-15$\,Myr and $\ds\sim1.0\pm0.2$\,kpc (\citealt{Pi5}, and 
references therein). In paper\,I we had assumed a constant SFR, uniform DR and a fixed 
binary fraction. Despite the latter two effects, the main source of differences is the 
SFR. In a constant SFR, stars of any age have the same probability of being formed, and 
young stars are significantly brighter than their older counterparts. Thus, when trying 
to match the same observed CMD, a simulation that contains an enhanced population of 
young stars (constant SFR) would require higher values of distance modulus than another 
based on a decaying SFR. At the same time, the constant SFR simulation would also imply 
a younger age.

\begin{figure}
\resizebox{\hsize}{!}{\includegraphics{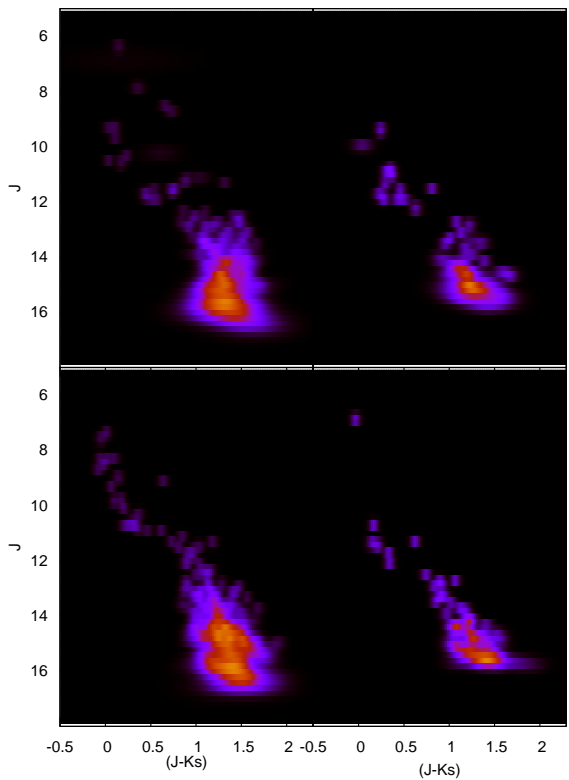}}
\caption[]{Observed (top panels) and simulated (bottom) Hess diagrams of Collinder\,197 
(left) and Pismis\,5 (right). Lighter grey shades indicate higher densities of stars. 
The simulated diagrams have been built with the mean parameters and the normal DR mode 
(Table~\ref{tab2}).}
\label{fig5}
\end{figure}

The representativeness reached by the minimisation process described above for 
Collinder\,197 and Pismis\,5 can be visually appreciated by comparing the observed 
and simulated Hess diagrams (Fig.~\ref{fig5}). The latter have been constructed with 
the mean parameters and assuming the normally-distributed DR
(Table~\ref{tab2}). Although the occurrence of some discreteness in both Hess diagrams, 
which is a natural consequence of the relatively low-mass nature of the clusters, 
especially for Pismis\,5, observation and simulation show a reasonable correspondence 
in both clusters.

\begin{figure}
\resizebox{\hsize}{!}{\includegraphics{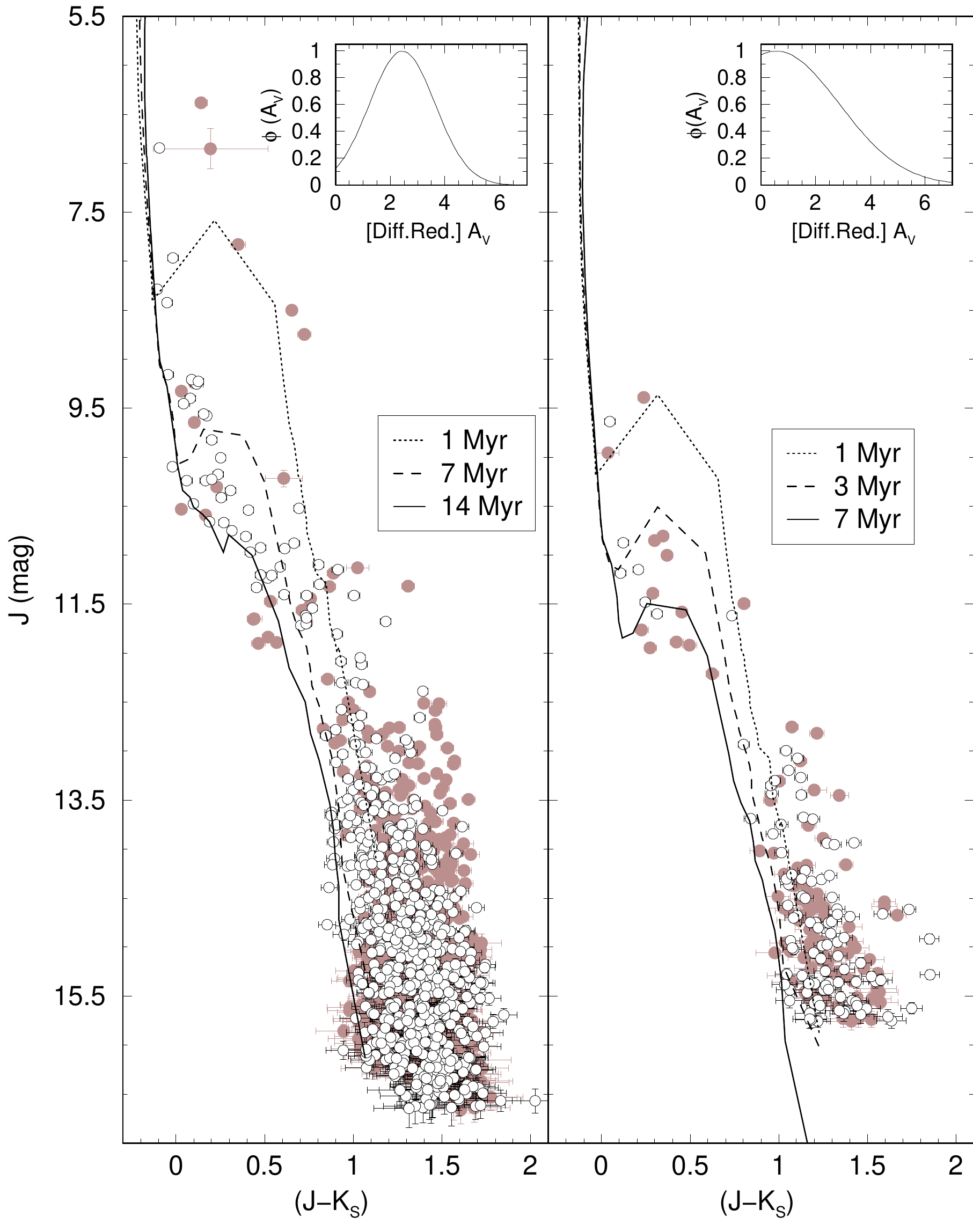}}
\caption[]{Field-star decontaminated CMDs (light-shaded circles) of Collinder\,197 (left) 
and Pismis\,5 (right) compared to a single realisation taken from the respective simulations 
(empty circles). The extraction radii are 10\arcmin\ (Collinder\,197) and 6\arcmin\ (Pismis\,5). 
Isochrones representing the star-formation history have been set according to the mean 
parameters (Table~\ref{tab2}). Insets show the DR distribution.}
\label{fig6}
\end{figure}

Finally, in Fig.~\ref{fig6} we compare a single CMD realisation randomly selected 
among the \nS\ simulated clusters - but having exactly the same number of stars - 
with the observed CMDs. The DR distributions (Collinder\,197 has higher \mdV\ and 
lower \sdr\ than Pismis\,5) are also shown. For illustrative purposes, Fig.~\ref{fig6} 
also shows the isochrones that represent the full star-formation history of both 
clusters. They have been set with the mean values of distance modulus and foreground 
reddening in Table~\ref{tab2}. Note that the optimum simulation and respective 
isochrone solution end up naturally respecting the blue border of the observed 
stellar sequences as a (not-imposed) boundary condition\footnote{From our perspective, 
this is somehow reassuring, since the blue border has been taken as a constraint to 
estimate fundamental parameters of young clusters with simpler methods such as that in 
\citet{vdB92}.}. Also, the fading and reddening effect of DR on the stars is clearly 
seen when one compares the youngest (and reddest) isochrone with the redwards spread 
of the PMS stellar sequences. This also shows that, if DR is not properly taken into 
account, fitting isochrones to a CMD would require somewhat younger ages coupled to 
higher values of distance modulus and foreground reddening, especially to account 
for the faint and red PMS stars together with the blue border.

Again, given that both are low-mass, young clusters, some morphological 
differences should be statistically expected, especially in the MS. Nevertheless, 
simulated and observed CMDs are similar in both cases.

\subsection{The low-mass stochasticity}
\label{RnS}

An interesting issue that can be investigated with \rr\ is the natural stochasticity 
associated with low-mass clusters. In other words, how the number of simulated clusters 
(\nS) affect the convergence pattern of the retrieved parameters. Obviously, \nS\ is 
expected to play an important role especially on a poorly-populated ($\Mcl<150\,\ms$) 
and young ($\ta\sim7$\,Myr) cluster such as Pismis\,5, in which the stochasticity tends 
to be critical. 

For a low-mass cluster consisting essentially of PMS stars, the stochasticity issue 
can be summarised as follows. Statistically, a random simulation of this cluster (same 
age, mass, etc) may contain a massive, bright star that is not present in the actual 
cluster. Consequently, because of the mass constraint, the CMD of this particular 
simulation would also lack a large fraction of the low-mass content. Thus, despite 
having been built with exactly the same parameters as the cluster, the \x2\ residuals 
of this simulation would be large, with a low representativeness. Given the extremely 
large number of parameter combinations, the probability of finding a single simulation 
with a Hess diagram matching that of the cluster is vanishingly low. On the other hand, 
when many simulations with exactly the same parameters are considered, the individual
Hess diagrams would certainly present different representativeness. Thus, the average 
Hess diagram over many simulations is expected to present a higher similarity (i.e., 
low \x2) with that of a cluster's than what would be obtained with a single, random 
simulation. It is in this context that averaging out several simulations of such twin 
clusters becomes an effective way of minimising stochasticity. 

\begin{figure}
\resizebox{\hsize}{!}{\includegraphics{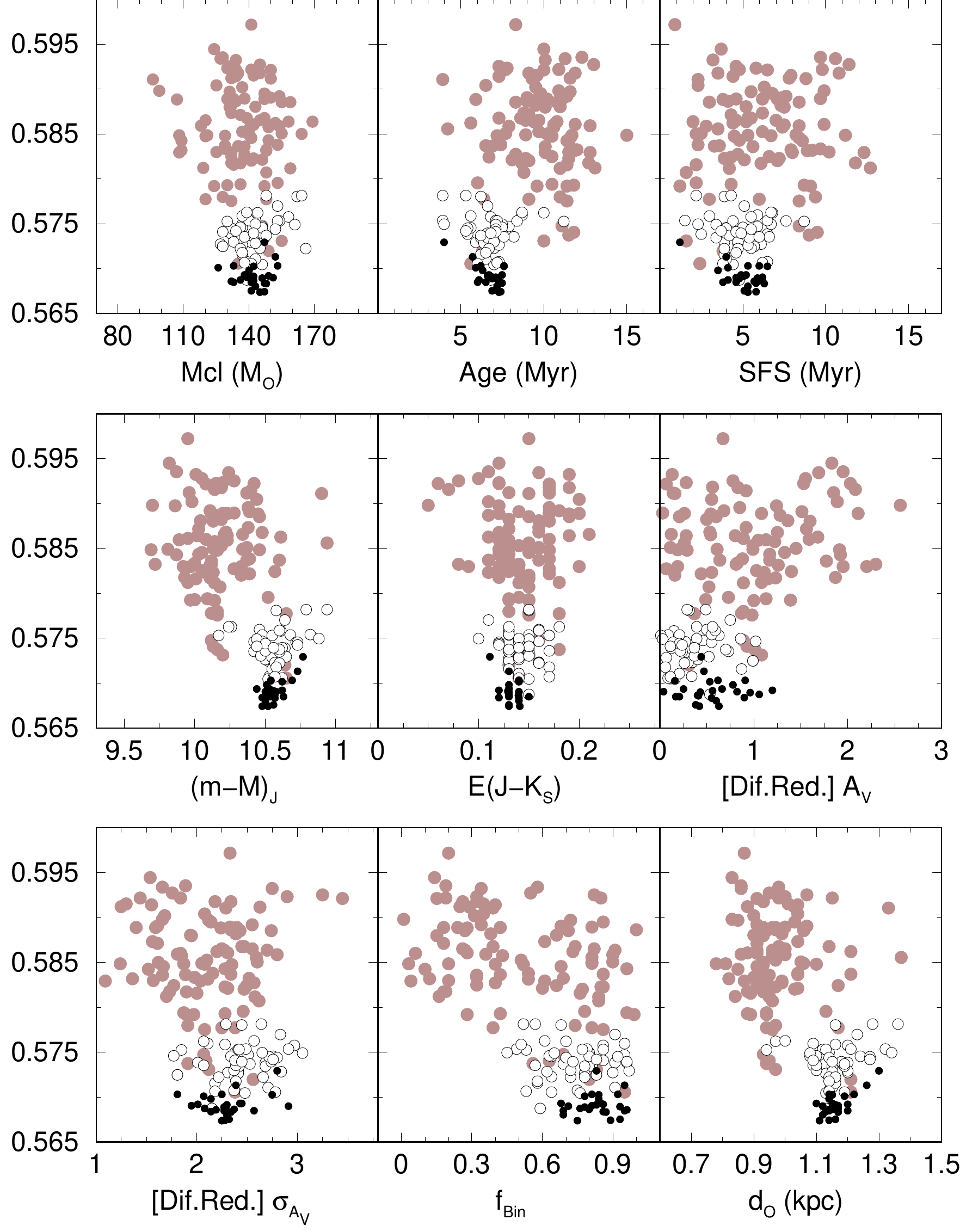}}
\caption[]{Increasing the number of twin simulations (\nS) improves the convergence 
pattern for a poorly-populated cluster such as Pismis\,5. Simulations built assuming
the normal DR mode. The number of simulations used in the runs are 
$\nS=10$ (light-shaded circles), $\nS=100$ (empty), and $\nS=500$ (filled).}
\label{fig7}
\end{figure}

Such effect is illustrated in Fig.~\ref{fig7}, in which we show the retrieved parameters
for Pismis\,5 after running \rr\ with $\nS=10,~100$ and $500$, always assuming the 
same initial conditions. Clearly, the convergence pattern, which is very loose for $\nS=10$, 
becomes tighter as \nS\ increases. Despite the significant scatter associated mainly with 
$\nS=10$, it is interesting to note that at the lowest \x2\ values, the parameters of 
$\nS=10$ and $100$ converge to those of $\nS=500$. Besides, the differences between the 
$\nS=100$ and $\nS=500$ runs are significantly smaller than those between $\nS=10$ and 
$\nS=100$.

Note, however, that some scatter is still present for SFS, mean DR and dispersion, and 
binary fraction, even when $\nS=500$. Ideally, a very large \nS\ should be used to reach 
a very tight convergence pattern for a low-mass cluster. In practice, however, this may 
lead to exceedingly long runtimes. In this sense, our recipe of using a moderate \nS\ 
combined to a series of runs may be taken as a compromise between runtime and robustness.

\subsection{\rr\ and the \x2\ topology}
\label{Min}

\begin{figure}
\resizebox{\hsize}{!}{\includegraphics{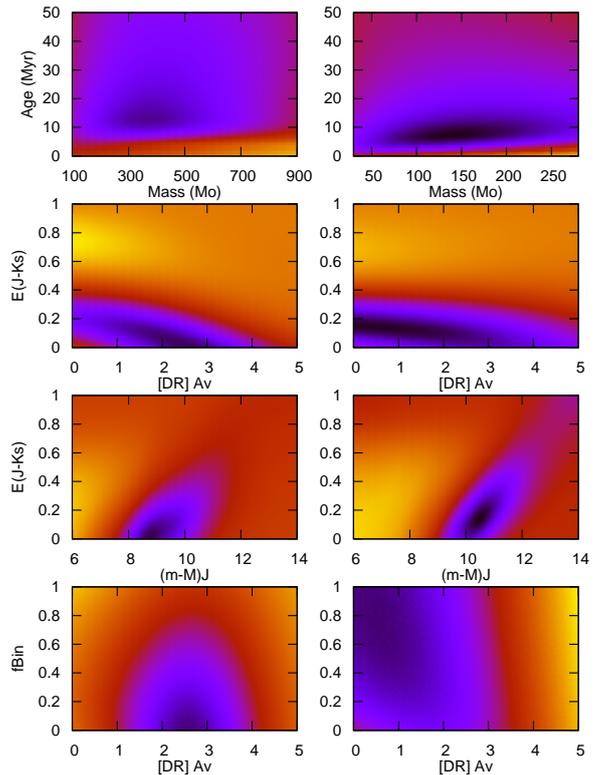}}
\caption[]{Selected two-dimensional \x2\ projections for Collinder\,197 (left) and 
Pismis\,5 (right). Darker colours indicate lower \x2. Convergence patterns vary 
significantly among different parameters.}
\label{fig8}
\end{figure}

Having found the best-fit parameters, we now use them to examine the topology 
of the \x2\ hyper-surface, something we address by means of selected two-dimensional 
projections (Fig.~\ref{fig8}). For statistical significance, the maps were produced
with $\nS=100$ and $500$ and the absolute minimum \x2\ parameters, respectively for 
Collinder\,197 and Pismis\,5. The presence of a minimum is clear, but with convergence 
patterns varying significantly among the parameters, being tight for most but somewhat 
loose especially for the binary fraction. 

Interestingly, the projections now are significantly smoother and with less features than 
the equivalent ones shown in Fig.~5 of Paper\,I. Possible reasons for such a contrast are 
that in Paper\,I the \x2\ hyper-surface {\em (i)} was built using some fixed parameters 
(binary fraction and SFS) and restrictive conditions (flat SFR and uniform DR distribution), 
{\em (ii)} the free parameters were allowed to vary within bins of fixed size (and not 
continuously as in the present approach) and within restricted ranges, and {\em (iii)} we 
used a different statistics for finding solutions. Apparently, the additional free parameters 
and less-restrictive conditions have raised the degeneracy of solutions found in Paper\,I. 

\begin{figure}
\resizebox{\hsize}{!}{\includegraphics{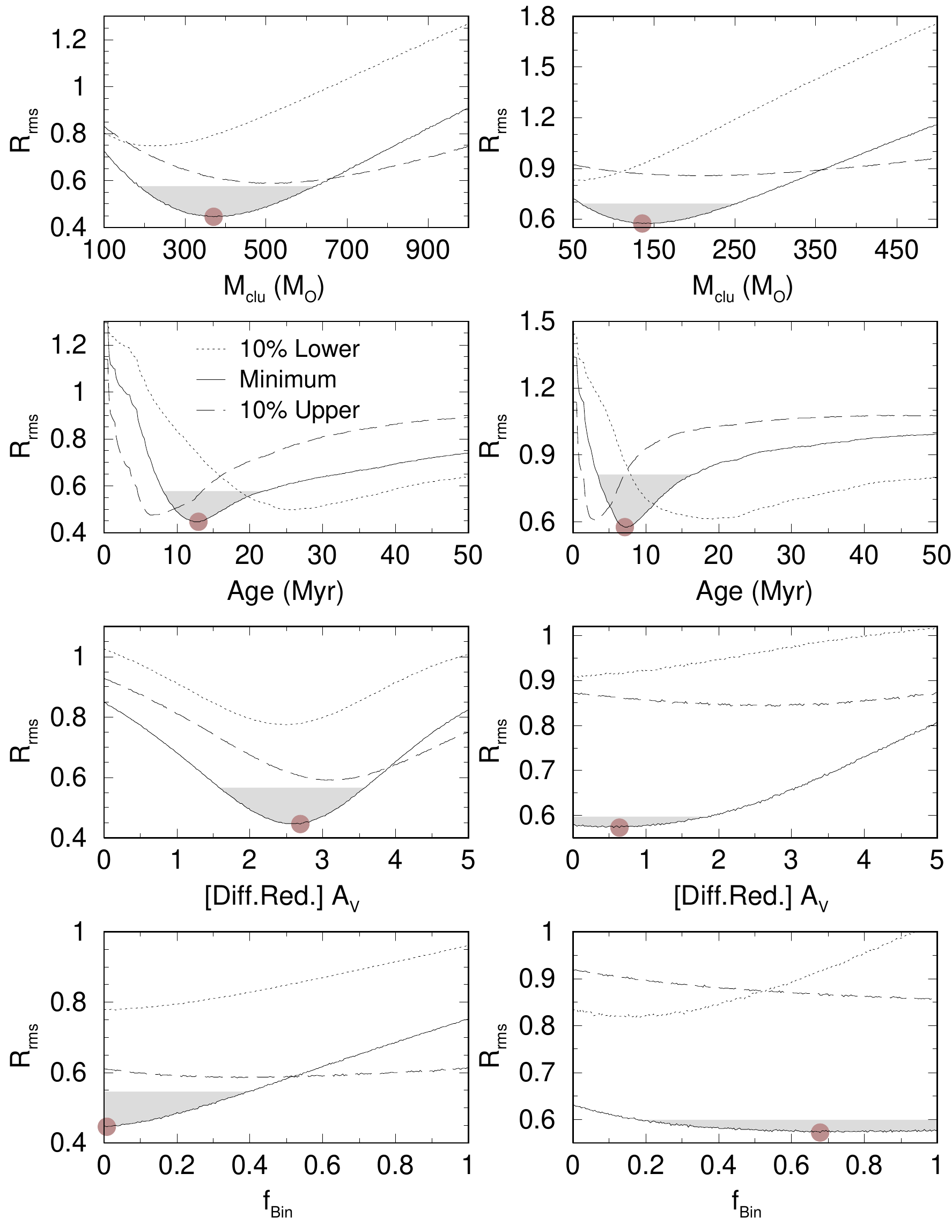}}
\caption[]{Selected one-dimensional projections of \x2\ computed with $\nS=500$ for 
Collinder\,197 (left) and $\nS=1500$ for Pismis\,5 (right). Besides the values computed 
for the absolute minimum parameters (solid line), we also show the \x2\ projections using 
values 10\% lower (dotted) and higher (dashed) than the minimum. The absolute minimum is 
shown by the shaded circle, and the approximate $1\sigma$ domain is shown by the grey region.}
\label{fig9}
\end{figure}

In any case, what is really important for our approach is the existence of at least one 
conspicuous minimum, and our expectation (Sect.~\ref{ASA}) is that ASA is efficient in 
finding its way towards the absolute minimum. In addition, a single minimum also serves 
to strengthen the unicity of the solutions. Thus, given the relevance of this assumption 
to the results discussed above, we summarise it in Fig.~\ref{fig9} with one-dimensional 
projections of the \x2\ hyper-surface for selected parameters of both clusters. For instance, 
we take from Table~\ref{tab2} the minimum values (obtained with the normal DR mode) for 
all the parameters except cluster mass, which is allowed to vary within a wide range (in 
this case, $0.5\le\Mcl(\ms)\le250$). Then, we compute \x2\ for masses within the adopted 
range, but keeping the remaining parameters fixed. For a deeper perspective on the effect 
of parameter variation on the \x2\ shape, this step is repeated with the fixed parameters 
changed to 10\% higher and lower than the optimum values. The same procedure is applied 
to the age, mean DR and binary fraction.

Consistently with the \x2\ maps (Fig.~\ref{fig8}), the selected \x2\ projections 
present a single minimum with a degree of definition (depth and width) that varies 
significantly among the adopted thresholds. When comparing different parameters, 
the minima tend to be quite narrow and deep for the age, somewhat wide and shallow 
for the binary fraction, and intermediate for the remaining parameters. 

The general conclusions emerging from the above discussion are: {\em (i)} the \x2\ 
hyper-surface contains at least one minimum within the adopted search range, {\em (ii)}
the morphological features of the minimum vary according to each parameter, and most 
importantly, {\em (iii)} \rr\ finds its way through the \x2\ topology towards the 
absolute minimum.

\section{Summary and conclusions}
\label{Conclu}

A new approach (\rr), designed to obtain a set of important parameters of young star 
clusters at a statistically significant confidence level, is presented in this paper. 
In short, it is essentially based on photometric properties and involves building 
realistic simulations of the Hess diagram of an actual star cluster, from which the 
residuals (\x2) with respect to the observed Hess diagram are computed. Besides 
cluster mass, age, foreground reddening and distance modulus, the simulations include 
the SFS, (alternative modes of) DR and the binary fraction as free parameters. The 
CMD spread due to photometric uncertainties is explicitly taken into account. 
Important features of the simulations are a linearly decreasing SFR and a normally 
distributed DR.

To find the absolute minimum of the \x2\ hyper-surface we use the global optimisation 
method known as adaptive simulated annealing (ASA), which is rather efficient and capable of 
escaping from local depressions. Given the highly-statistical nature of the simulations (and 
the cluster stochasticity), we show that an acceptable parameter-retrieval rate is achieved 
by combining a moderate number of simulated clusters with a series of independent ASA runs,
while realistic errors in the derived parameters are obtained by exploring properties of the 
depression shape. Tests with model clusters built with a broad range of parameters show that 
the distribution of retrieved values (corresponding to the absolute \x2\ minimum) usually 
follows a convergence pattern that is tighter as \x2\ declines, with the same occurring with 
actual clusters. We also find that the parameter retrieval presents a high sensitivity for 
cluster mass, distance modulus and foreground reddening, but dropping somewhat for the remaining 
parameters. 

We remark that the particular results discussed in this work may be somewhat model dependent, 
in the sense that they are based on the 2MASS near-infrared photometry coupled to the Padova 
and Siess MS and PMS isochrone sets. Other isochrones with somewhat different mass to light
ratios for individual stars may possibly affect the star cluster parameters when retrieved by 
\rr, especially the age, mass and distance. In any case, \rr\ can be easily adapted to any 
photometric system (and isochrone set), provided the respective MS and PMS isochrones and the 
relation of photometric errors with magnitude are available. However, by allowing a deeper view 
through the dust, the near-infrared seems to be the best window to disentangle the effects of 
DR, binaries and SFS.

In summary, we show in this work that our semi-analytical and comprehensive approach is 
capable of uncovering a series of parameters of young clusters, even when photometry is 
the only available information. Among these, the cluster age, star-formation spread, mass 
and binary fraction are important for establishing the dynamical state of a cluster, or 
derive a more precise SFR in the Galaxy.

\section*{Acknowledgements}
We thank the referee, Melvin Hoare, for important comments and suggestions. 
This publication makes use of data products from the Two Micron All Sky Survey, which is a
joint project of the University of Massachusetts and the Infrared Processing and Analysis
Center/California Institute of Technology, funded by the National Aeronautics and Space
Administration and the National Science Foundation. We acknowledge financial support from 
the Brazilian Institution CNPq.

\label{lastpage}
\end{document}